\documentclass[reprint,amsmath,amssymb,aps,pra,floatfix,longbibliography]{revtex4-2}
\usepackage[T1]{fontenc}
\usepackage{lmodern,microtype}
\usepackage{braket,graphicx,bm}
\usepackage{xcolor}
\definecolor{docnotelinkcolor}{rgb}{0,0,0}
\usepackage{tikz,booktabs}
\usetikzlibrary{arrows.meta,decorations.pathreplacing}

\usepackage{orcidlink}

\newtheorem{proposition}{Proposition}
\makeatletter
\AtBeginDocument{\immediate\write\@auxout{\string\citation{paperBibliographyStyle}}}
\makeatother
\begin{document}
\raggedbottom
\title{Exact entanglement trade-offs in qutrit and composite-dimensional stabilizer states}
\author{Zakaria Dahbi\orcidlink{0000-0001-9933-2184}}
\email{zakaria.dahbi@kcl.ac.uk}
\affiliation{Attosecond Quantum Physics Laboratory, Department of Physics, King's College London, London WC2R 2LS, United Kingdom}
\date{September 29, 2026}
\begin{abstract}
When absolute maximal entanglement is unavailable in a state family, different objectives can select different distributions of residual entanglement. We determine exact trade-offs for qutrit stabilizer states and binary--ternary product states. For eight qutrits, evaluation of the complete published classification gives three nondominated balanced-cut entanglement profiles, characterized by the probabilities of obtaining three or four maximally entangled qutrit pairs. We relate the squared entropy-deficit objective to code weight enumerators and use a uniformity-forcing bound to prove that its eleven-qutrit minimum is six. The unique minimizing Clifford class is an already classified code, whose entropy profile also attains a sharp bound on average reference--erasure mutual information among four-uniform eleven-qutrit stabilizer states. In local dimension six, explicit sector alignments attain 56 maximally mixed four-party reductions, the maximum for arbitrary binary--ternary product states. Within the stabilizer product family, the exact cost--count Pareto boundary consists of $(44,48)$ and $(48,56)$. These results identify when aggregate optimization determines operational entanglement properties and when it does not. Published classifications supply coverage; analytical bounds, explicit witnesses, and independent arithmetic checks establish the stated consequences.
\end{abstract}
\maketitle
\section{Introduction}
Absolutely maximally entangled (AME) states have maximally mixed reductions on every subsystem containing no more than half their parties \cite{scott2004multipartite,helwig2012absolute,goyeneche2015absolutely}. Their connections to quantum error correction, secret sharing, and perfect tensors make both constructions and obstructions useful \cite{knill1997theory,rains2002nonbinary,raissi2018optimal,alsina2021absolutely,cleve1999share,mansour2020quantum,pastawski2015holographic}. When an ansatz cannot contain an AME state, however, a nonexistence statement does not determine how small its residual entropy losses can be, or how those losses are distributed among subsystems.

For graph states, the rank of the adjacency block crossing a bipartition determines its entanglement spectrum \cite{hein2004multiparty,marin2013access,helwig2013absolutely,fattal2004entanglement}. Finite-field graph-state and symmetric-matrix constructions are established \cite{chen2016graph,feng2017multipartite}; the cut rank here is distinct from general multipartite tensor rank \cite{oum2017rank,horodecki2009quantum,bruzda2019tensor}. Earlier searches optimized marginal purity and other multiple-cut criteria \cite{facchi2008maximally,facchi2009statistical,brown2005searching,reuvers2018algorithm}. We use this existing entanglement description to evaluate an integer-valued sum of squared entropy deficits. Its minimum is zero exactly at an AME construction in the chosen family, while a strictly positive minimum quantifies an obstruction for the specified objective.

We ask which entanglement distributions remain optimal when the performance criterion changes. The squared cut-rank deficit supplies a tractable scalar objective, but the physical questions concern the probabilities of obtaining a specified entanglement yield, the number of maximally mixed reductions, and information leaked by erasure. We determine complete finite-family trade-offs for these quantities rather than interpreting a scalar optimum as universally optimal entanglement.

Three results provide the main applications. First, the 817 published eight-qutrit classes yield exactly three nondominated balanced-cut rank distributions. Second, a uniformity-forcing bound and Danielsen's published length-eleven weight enumerators give the exact eleven-qutrit squared-deficit minimum six and a sharp average erasure-leakage bound. Third, aligning eight-party binary and ternary sectors gives 56 maximally mixed four-party reductions in local dimension six, saturating a bound valid for arbitrary pure states that factor across those sectors. Within the stabilizer product family, we also determine the exact trade-off between the squared cost and that count.

The entanglement--rank relation, graph-state representations, and code classifications are established inputs \cite{hein2004multiparty,fattal2004entanglement,danielsen2009graph,danielsen2012classification}. Cabello \emph{et al.} already tabulated eight-qubit rank profiles \cite{cabello2009eight,cabello2010erratum}; Zhang \emph{et al.} proved that an arbitrary eight-qubit pure state has at most 56 maximally mixed four-party reductions \cite{zhang2025extremal}. Our contributions are the objective-dependent trade-offs, the certification argument for eleven qutrits, and their explicitly scoped information-theoretic consequences. We do not claim new stabilizer codes or unrestricted AME obstructions. AfAME provides a search and verification implementation: when zero-cost AME construction is unavailable in a chosen family, it can construct and, when an optimum is certified, verify an objective-specific maximally multipartite entangled state (MMES) for the declared cost or count criterion.

\section{Quadratic-phase states and the optimization objective}
\subsection{Finite-field states and cut spectra}
All states considered here are pure. Let $\mathbb F=\mathbb F_{p^m}$, with $q=|\mathbb F|$, and let $\chi(x)=\exp[2\pi i\operatorname{Tr}_{\mathbb F/\mathbb F_p}(x)/p]$ be its canonical additive character \cite{lidl1997finite}. For a symmetric $N\times N$ matrix $P$ over $\mathbb F$, define $\phi_P(\mathbf x)=\sum_iP_{ii}x_i^2+\sum_{i<j}P_{ij}x_ix_j$ and the normalized phase state
\begin{equation}
\ket{\Phi_P}=q^{-N/2}\sum_{\mathbf x\in\mathbb F^N}\chi\!\left(\phi_P(\mathbf x)\right)\ket{\mathbf x}.
\label{eq:phase}
\end{equation}
For $m=1$, these are weighted graph states up to local diagonal phases. The family is a restricted class of stabilizer states; the diagonal of $P$ contributes only local phases and cannot change cut entanglement \cite{helwig2013absolutely,gottesman1997stabilizer,dehaene2003clifford,feng2017multipartite}.

For a subsystem $S$, let $r_S=\operatorname{rk}_{\mathbb F}P_{S,\bar S}$. The established cut-rank relation gives $q^{r_S}$ nonzero eigenvalues of $\rho_S$, each equal to $q^{-r_S}$ \cite{hein2004multiparty,marin2013access,fattal2004entanglement}. Hence its purity and von Neumann and second R\'enyi entropies are
\begin{equation}
\operatorname{Tr}(\rho_S^2)=q^{-r_S},\qquad S(\rho_S)=S_2(\rho_S)=r_S\ln q.
\label{eq:rpd}
\end{equation}
Appendix 1 derives the complete spectrum directly from the amplitudes in Eq.~\eqref{eq:phase}, without invoking stabilizer generators. Equation~\eqref{eq:rpd} implies the familiar AME criterion within this family: every cut with $|S|\le\lfloor N/2\rfloor$ must satisfy $r_S=|S|$. Checking all cuts at half size suffices for certification, since maximally mixed larger reductions have maximally mixed subreductions. We nevertheless score all nonempty cuts up to half size to show where imperfect candidates lose entropy.

\subsection{An integer objective with an exact zero set}
\label{sec:optimization}
Let $\mathcal{S}_N=\{S\subseteq[N]:1\leq |S|\leq\lfloor N/2\rfloor\}$.
For a symmetric phase matrix $P$ over $\mathbb{F}_q$, define the nonnegative integer
\begin{equation}
 \delta_S(P)=|S|-\operatorname{rk}_{\mathbb{F}_q}(P_{S,\bar S}).
 \label{eq:deficit}
\end{equation}
Equation~\eqref{eq:rpd} gives $\Delta_S=|S|\ln q-S(\rho_S)=\delta_S\ln q$. We therefore optimize
\begin{equation}
\mathcal{C}_q(P)=\sum_{S\in\mathcal{S}_N}\delta_S(P)^2
=\frac{1}{(\ln q)^2}\sum_{S\in\mathcal{S}_N}\Delta_S^2 .
\label{eq:cost}
\end{equation}
This is an exact reformulation of AME construction \emph{within the ansatz}: $\min_P\mathcal{C}_q(P)=0$ if and only if the ansatz contains an $\operatorname{AME}(N,q)$ state. A matrix with cost zero is a certificate once every cut rank has been recomputed exactly. At positive cost the objective distinguishes a single large rank loss from several smaller losses. It is a sum of squared entropy shortfalls for one field, not a distance in Hilbert space from the set of all AME states.

When the minimum is positive, we use MMES in the objective-specific sense: a state attaining the minimum of the declared cost, or the maximum of a declared operational count such as $F$. This terminology does not assert that one scalar objective defines a universally preferred MMES, and a heuristic search output is not by itself an optimality certificate.

The entropy shortfall also has a direct information-theoretic interpretation. With $\tau_S=I/q^{|S|}$ and $D(\rho\Vert\sigma)=\operatorname{Tr}[\rho(\ln\rho-\ln\sigma)]$,
\begin{equation}
D(\rho_S\Vert\tau_S)=|S|\ln q-S(\rho_S)=\delta_S\ln q.
\label{eq:relative_entropy}
\end{equation}
Thus Eq.~\eqref{eq:cost} sums the squared relative entropies of the marginals from their maximally mixed references, in units of $(\ln q)^2$. If $M=|\mathcal S_N|$, then $\mathcal C_q/M=\overline\delta^{\,2}+\operatorname{Var}(\delta)$ for the uniform distribution over the scored subsets. This identity displays the simultaneous penalty on mean deficit and its variation across cuts; it does not assert that the squared penalty is the unique physically preferred objective.

For pairwise coprime prime-power field sectors $q_\alpha=p_\alpha^{m_\alpha}$ with $d=\prod_\alpha q_\alpha$, AfAME uses a tuple of symmetric matrices $\mathbf P=(P^{(1)},\ldots,P^{(r)})$ and the sector-resolved objective
\begin{equation}
\mathcal{C}_d(\mathbf P)
=\sum_{S\in\mathcal{S}_N}\sum_{\alpha=1}^{r}
\left(|S|-\operatorname{rk}_{\mathbb{F}_{q_\alpha}}
P^{(\alpha)}_{S,\bar S}\right)^2 .
\label{eq:composite_cost}
\end{equation}
The shorthand $\operatorname{rk}_{\mathbb F_{q_\alpha}}P^{(\alpha)}_{S,\bar S}$ means the rank of that cut matrix. Zero cost is equivalent to every sector attaining maximal entropy on every required subset. The total entropy shortfall of a cut is $\sum_\alpha\delta_{S,\alpha}\ln q_\alpha$, but its square is \emph{not} the objective in Eq.~\eqref{eq:composite_cost}; the latter retains separate field-sector information. For even $N$ the set $\mathcal{S}_N$ contains both members of each balanced complementary pair, as does the implementation. This duplication leaves the zero set unchanged and weights balanced cuts twice.

Write $\mathcal C_q^*(N)$ for the minimum of Eq.~\eqref{eq:cost} over symmetric $N\times N$ matrices over $\mathbb F_q$, and analogously define $\mathcal C_d^*(N)$ from Eq.~\eqref{eq:composite_cost}. Since the sector matrices are independent variables, their optimal costs separate exactly:
\begin{equation}
\mathcal C_d^*(N)
=\sum_{\alpha=1}^{r}\mathcal C_{q_\alpha}^*(N).
\label{eq:min_separates}
\end{equation}
Each summand depends on only one matrix, so choosing a minimizer of each summand attains the sum of the lower bounds. Equation~\eqref{eq:min_separates} applies to the stated product ansatz.

Define $F(P)=|\{S\in\mathcal S_N:\delta_S(P)>0\}|$. Because every nonzero deficit is an integer, $F(P)\leq\mathcal C_q(P)$; the cost also bounds the squared loss on any individual cut. In the product case the corresponding inequality holds for the number of failing cut-sector incidences. These relations make a nonzero search result informative, but a heuristic search can only supply an upper bound on the minimum cost. Independent theorems or exhaustive enumeration are needed to establish a positive lower bound. Earlier searches have minimized subsystem purities or other entanglement objectives \cite{facchi2008maximally,facchi2009statistical,brown2005searching,reuvers2018algorithm}. Our choice exploits the integral rank-deficit landscape of phase matrices.

\section{Certification from uniformity and weight enumerators}
\label{sec:uniformity}
A low deficit cost can force exact uniformity, thereby reducing an optimization over pure stabilizer states to a published classification at a fixed minimum distance. For prime local dimension, every pure stabilizer state is locally Clifford equivalent to a weighted graph state; we extend $\delta_S=|S|-S(\rho_S)/\ln q$ and $\mathcal C_q=\sum_{S\in\mathcal S_N}\delta_S^2$ to these states. Their marginal spectra, integer deficits, and costs are preserved by this equivalence.
\begin{proposition}
If a $t$-party marginal, $1\le t\le k$, of a pure prime-dimensional stabilizer state has deficit at least $a\ge1$, then, with $k=\lfloor N/2\rfloor$,
\begin{equation}
\mathcal C_q\ge a^2\sum_{j=t}^{k}\binom{N-t}{j-t}.
\label{eq:uniformity_bound}
\end{equation}
In particular, $\mathcal C_q<N-k+2$ implies $(k-1)$-uniformity.
\end{proposition}
\noindent\emph{Proof.} For $T\subseteq U$, subadditivity gives $S(\rho_U)\le S(\rho_T)+(|U|-|T|)\ln q$, hence $\delta_U\ge\delta_T$. Every scored superset of $T$ contributes at least $a^2$. If a state is not $(k-1)$-uniform, choose a deficient $(k-1)$-subset; it and its $N-k+1$ supersets of size $k$ give the last assertion. Integer deficits are essential for this threshold.\hfill$\square$

For a $(k-1)$-uniform qutrit stabilizer state, every $k$-party deficit belongs to $\{0,1,2\}$. Indeed, $S(\rho_S)\ge S(\rho_T)-\ln3=(k-2)\ln3$ for $T\subset S$ of size $k-1$. Write $A_k$ for the number of weight-$k$ stabilizer elements, counting each phase-fixed group element once. The subgroup supported within $S$ contains $3^{\delta_S}$ elements. All its nonidentity elements have weight exactly $k$, so
\begin{equation}
 A_k=\sum_{|S|=k}(3^{\delta_S}-1)
 =2\sum_{|S|=k}\delta_S^2=2\mathcal C_3.
 \label{eq:weight_cost}
\end{equation}
The last equality uses $\delta^2=(3^\delta-1)/2$ for $\delta=0,1,2$, and smaller marginals contribute zero. This is a conditional identity, not a formula for arbitrary stabilizer states.

The connection also locates the objective relative to Facchi's balanced-purity potential $\pi_{\rm ME}=\binom Nk^{-1}\sum_{|S|=k}\operatorname{Tr}\rho_S^2$ \cite{facchi2008maximally}. On this same qutrit subclass,
\begin{equation}
 \mathcal C_3=\frac{\binom Nk}{2}\bigl(3^k\pi_{\rm ME}-1\bigr).
 \label{eq:facchi_equivalence}
\end{equation}
Thus the two objectives have identical minimizers there. They differ in general: for graph states, balanced purity weights a deficit by $q^\delta$, whereas Eq.~\eqref{eq:cost} weights it quadratically and also includes smaller subsets. Facchi \emph{et al.} additionally considered mean purity plus a tunable standard-deviation penalty; distributional uniformity is already part of that framework.

\section{Exact eight-party minima}
\label{sec:exact8}
\begin{proposition}
For the finite-field quadratic-phase family and the cut-counting convention of Eqs.~\eqref{eq:cost} and \eqref{eq:composite_cost},
\begin{equation}
\mathcal C_2^*(8)=28,\qquad \mathcal C_3^*(8)=16,
\label{eq:prime8_minima}
\end{equation}
and for the binary--ternary product family,
\begin{equation}
\mathcal C_6^*(8)=44.
\label{eq:eight_six_bounds}
\end{equation}
\end{proposition}

\noindent\emph{Proof by exhaustive evaluation.}
The diagonal of a phase matrix does not enter any cut block. Every symmetric binary matrix is therefore represented, for this objective, by a simple graph. Starting with all 1044 nonisomorphic seven-vertex graphs \cite{read1998atlas} and attaching an eighth vertex in all $2^7$ ways covers every eight-vertex graph up to relabelling. Exact elimination and a separate row-span calculation both give minimum 28 over the resulting 133,632 extensions. A witness has 140 full-rank cuts, 20 cuts of deficit one, and two of deficit two, giving $20+2\times4=28$ (Appendix 4).

For the ternary sector, every qutrit stabilizer state has a weighted graph representative under partywise Clifford transformations. The published classification of self-dual additive codes over $\mathbb F_9$ contains 817 length-eight classes: 659 indecomposable and 158 decomposable \cite{danielsen2009graph,danielsen2012classification,danielsenDatabase}. These equivalences preserve subsystem entropies; coordinate permutations merely permute the subsets in the sum. Consequently, evaluating one graph representative per class suffices to minimize $\mathcal C_3$ over all symmetric ternary phase matrices. This coverage also extends the same entropy-deficit minimum to all eight-qutrit stabilizer states.

We evaluate all 162 required subsets of every representative, giving 132,354 cut ranks. One calculation uses Gaussian elimination over $\mathbb F_3$. An independent calculation enumerates the $3^8$ stabilizer labels of each graph and counts the elements supported within each subset, without using elimination. The two complete cut-rank lists agree. The minimum is 16, attained by ten classes; a witness has 54 full-rank balanced cuts and 16 balanced cuts of deficit one, with all smaller cuts full rank. Appendix 5 gives the matrix, independent check, and provenance. Finally, the independent sector variables in Eq.~\eqref{eq:min_separates} give $\mathcal C_6^*(8)=28+16=44$. The supplied composite certificate attains these two sector costs.\hfill$\square$

The completeness inputs are the published graph atlas and code classification, not heuristic search outcomes. The new evaluation strengthens the ternary nonexistence statement into a sharp positive minimum for the stated objective. For comparison, excluding a ternary AME state alone gives only $\mathcal C_3^*(8)\ge2$, since a deficient balanced cut is accompanied by its complement. Together with the binary minimum this would give only $30\le\mathcal C_6^*(8)\le44$; the exhaustive ternary calculation closes that interval.

\subsection{A complete entanglement-yield trade-off}
All ten minimizing ternary classes have maximally mixed one-, two-, and three-party reductions. Their four-party deficit histograms fall into the three profiles in Table~\ref{tab:ternary_profiles}. Let $n_j$ count balanced subsets with $\delta_S=j$. Each profile satisfies $n_1+4n_2=16$, but the total deficit $n_1+2n_2$ and the number $n_0$ of maximally mixed reductions differ.

\begin{table*}[t]
\caption{Balanced-cut profiles of all ten minimizing eight-qutrit classes. Counts include both complementary four-party subsets. Every class has $\mathcal C_3=16$ and all 92 smaller reductions maximally mixed. $L=\sum_{|S|=4}\delta_S$ determines the mean balanced-cut entropy $[4-L/70]\ln3$.}
\label{tab:ternary_profiles}
\centering
\begin{tabular}{rrrrr}
\toprule
$n_0$ & $n_1$ & $n_2$ & Classes & $L$\\
\midrule
54 & 16 & 0 & 7 & 16\\
60 & 8 & 2 & 2 & 12\\
66 & 0 & 4 & 1 & 8\\
\bottomrule
\end{tabular}
\end{table*}

All ten states also share the weight distribution $(1,0,0,0,32,320,1088,2560,2560)$. Equation~\eqref{eq:weight_cost} therefore explains their equal cost and equal average balanced purity. Their cut-resolved distributions contain information absent from this common weight enumerator.

Consider a uniformly selected four-versus-four bipartition. The two blocks may each perform joint local operations. A cut of rank $r$ is locally unitarily equivalent across the blocks to $r$ maximally entangled qutrit pairs and unentangled ancillary systems, since its Schmidt spectrum is flat with $3^r$ entries. Without additional shared entanglement, local operations and classical communication between the blocks cannot increase Schmidt rank, even conditionally. Define $p_j=\Pr(r\ge j)$ for this task. These probabilities concern block-local operations, not operations separately restricted to the eight individual parties.

\begin{proposition}
Over all pure eight-qutrit stabilizer states, the complete nondominated rank-tail vectors $(70p_4,70p_3,70p_2,70p_1)$ are
\begin{equation}
 (54,70,70,70),\quad(60,68,70,70),\quad(66,66,70,70).
 \label{eq:tail_frontier}
\end{equation}
They are attained by the ten classes in Table~\ref{tab:ternary_profiles}. Every such state obeys
\begin{equation}
 p_4\le\frac{33}{35},\qquad p_4+3p_3\le\frac{132}{35}.
 \label{eq:tail_bound}
\end{equation}
\end{proposition}
\noindent\emph{Proof.} Evaluate the four integer rank-tail counts for each of the 817 classified representatives, and remove any vector componentwise dominated by another. Both independent complete cut-rank calculations give precisely Eq.~\eqref{eq:tail_frontier}. All vectors lie below at least one of these three; the bounds follow by direct substitution. Appendix 8 specifies the comparison and supplies every evaluated profile.\hfill$\square$

\begin{figure*}[tb]
\centering
\includegraphics[width=0.94\textwidth]{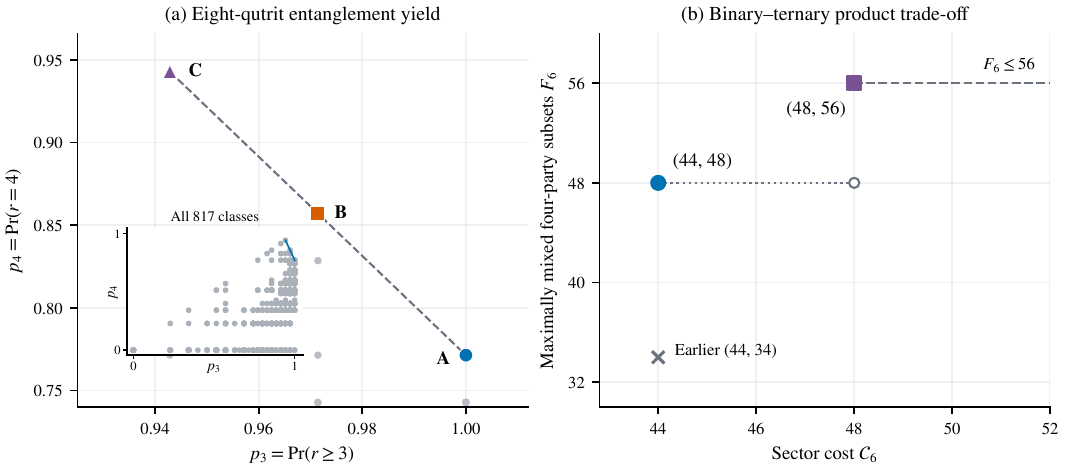}
\caption{Exact entanglement trade-offs. (a) Enlarged view of the eight-qutrit boundary; the inset shows all 817 classified states over the full probability range (coincident points overlap). The circle A, square B, and triangle C have $(70p_3,70p_4)=(70,54),(68,60),(66,66)$, respectively. Grey points are other classified profiles. The dashed segment is attainable by labelled randomization between A and C; it does not assert a continuum of pure stabilizer profiles. Operations are local to the two four-party blocks. (b) Filled markers are the two deterministic stabilizer-product Pareto points. The dotted segment gives the upper bound $F_6\le48$ for $44\le\mathcal C_6<48$; its open endpoint is excluded from that range. The dashed line gives $F_6\le56$, also valid for arbitrary pure binary--ternary product states. These lines indicate bounds, not additional attaining states. The earlier certificate is dominated at the same cost.}
\label{fig:tradeoffs}
\end{figure*}

This determines the optimum for any nondecreasing utility of the achievable pair number: at least one of these three profiles maximizes its expectation. The middle profile is the midpoint of the other two. Randomization between the endpoint resources, with the selected preparation known to the parties, realizes their connecting segment; such randomization is not assumed in the deterministic classification. The first endpoint guarantees three pairs on every cut, whereas the last maximizes the probability of four pairs. Its higher mean entropy comes with a two-pair yield on four cuts. The squared-deficit objective ties all three. Figure~\ref{fig:tradeoffs} shows this boundary.

\subsection{Comparison with eight-qubit purity optimization}
The binary minimizing profile $(48,20,2)$ is already present as class 146 in Cabello \emph{et al.}'s rank table, after doubling its balanced-bipartition counts \cite{cabello2009eight,cabello2010erratum}. It has $\pi_{\rm ME}=3/35$ and cost 28. Zhang \emph{et al.}'s graph $T_4$ has profile $(56,8,6)$, the same average purity, and cost 32 \cite{zhang2025extremal}. Thus the quadratic score distinguishes two states tied by bare mean purity. The $T_4$ state instead maximizes the number of fully mixed four-party reductions. This comparison is not a claim that the binary minimizing state is new, or that the quadratic objective improves on every purity-dispersion criterion.

\subsection{Alignment and the six-level cost--count boundary}
Let $F_6$ count maximally mixed four-party reductions of a state that, after identifying each local space with $\mathbb C^2\otimes\mathbb C^3$, factors as $|\psi_2\rangle\otimes|\psi_3\rangle$. A reduced product is maximally mixed exactly when both factors are maximally mixed, so its set of good subsets is the intersection of the two sector sets. The scalar cost ignores their relative alignment.

\begin{proposition}
For arbitrary pure binary--ternary product states on eight parties,
\begin{equation}
\max F_6=56.
\label{eq:product_full_bound}
\end{equation}
Within the stabilizer product family, the complete Pareto boundary for minimizing $\mathcal C_6$ and maximizing $F_6$ is
\begin{equation}
 (\mathcal C_6,F_6)=(44,48),\quad(48,56).
 \label{eq:product_pareto}
\end{equation}
\end{proposition}
\noindent\emph{Proof.} Zhang \emph{et al.}'s unrestricted eight-qubit bound gives $F_6\le F_2\le56$ \cite{zhang2025extremal}. For attainment, take their binary $T_4$ graph and align the four deficient subsets of ternary record 749 inside its 14 deficient subsets. The explicit permutation and matrices in Appendix 8 give $F_6=56$, with sector costs 32 and 16.

For the cost--count boundary, exhaustive binary evaluation gives $F_2\le48$ whenever $\mathcal C_2<32$, with the attainable nondominated pairs $(28,48)$ and $(32,56)$. Independent row-span and elimination-based checks cover all 133,632 graph extensions. Since $\mathcal C_3\ge16$, every product with $\mathcal C_6<48$ has $F_6\le48$. A second explicit alignment of record 749 with the cost-28 binary witness attains $(44,48)$. The lower cost bound 44 and upper count bound 56 then exclude every other Pareto point.\hfill$\square$

The earlier AfAME certificate had $(\mathcal C_6,F_6)=(44,34)$. It was optimal for the scalar score but dominated in the cost--count plane. The new alignment reaches 48 at the same cost; increasing the cost to 48 permits the maximal count 56. Equation~\eqref{eq:product_full_bound} holds beyond stabilizer sectors, but still requires the stated tensor factorization. It is not a bound on arbitrary eight-party six-level states.

\section{An exact eleven-qutrit optimum and erasure leakage}
\label{sec:eleven}
The uniformity threshold makes it possible to prove an optimum without classifying all length-eleven codes, whose classification would be much larger than the classification at optimal minimum distance.
\begin{proposition}
Over all pure eleven-qutrit stabilizer states,
\begin{equation}
 \mathcal C_3^*(11)=6.
 \label{eq:eleven_optimum}
\end{equation}
The minimizing class is unique under local Clifford operations and party permutations: it is Danielsen's published class with weight enumerator parameter $\alpha=0$ \cite{danielsen2012classification}.
\end{proposition}
\noindent\emph{Proof.} A state that is not four-uniform has a deficient four-party subset. That subset and its seven five-party supersets give $\mathcal C_3\ge8$ by Eq.~\eqref{eq:uniformity_bound}. For a four-uniform state, Eq.~\eqref{eq:weight_cost} gives $\mathcal C_3=A_5/2$. Danielsen classified all length-eleven codes of optimal minimum distance five and established the absence of a distance-six code; larger distances are excluded by the Singleton bound. The allowed weight-five coefficients are $A_5=12+2\alpha$, with $\alpha\in\{0,6,7,\ldots,50,54,60\}$; the $\alpha=0$ class is unique \cite{danielsen2012classification}. Hence every four-uniform state has cost at least six. Direct verification of the published matrix gives 456 full-rank five-party reductions and six of deficit one, with all 561 smaller reductions maximally mixed. This attains six and proves uniqueness in the stated equivalence relation.\hfill$\square$

The attaining state is established prior work, not a newly discovered code. The contribution here is the global objective certificate: uniformity propagation excludes every lower-distance competitor, so the published optimal-distance classification suffices. The previously reported AfAME candidate with cost 12 is a valid but suboptimal construction. Appendix 8 supplies the published attaining matrix, its exact cut list, and an independent enumeration of its $3^{11}$ stabilizer labels.

For an information-theoretic interpretation, choose any party $R$ of a four-uniform eleven-qutrit state as a reference. Its maximally mixed reduction makes the pure state the normalized Choi state of an isometric encoding of one qutrit into the other ten. If a four-party subset $E$ of those ten is erased, then
\begin{equation}
 I(R:E)=S(\rho_R)+S(\rho_E)-S(\rho_{RE})
 =\delta_{RE}\ln3.
 \label{eq:erasure_leakage}
\end{equation}
Zero mutual information is the decoupling condition for exact recovery from the remaining systems \cite{knill1997theory,cleve1999share}. This interpretation concerns erasure of known locations and this specified encoding.

Let the reference be uniformly chosen from the eleven parties and then choose $E$ uniformly among its $\binom{10}{4}=210$ four-party erasure patterns. Every five-subset occurs five times, giving
\begin{equation}
 \mathbb E_{R,E}I(R:E)
 =\frac{\ln3}{462}\sum_{|S|=5}\delta_S
 \ge\frac{\ln3}{77}.
 \label{eq:leakage_bound}
\end{equation}
To prove sharpness, the unique $\alpha=0$ state has total deficit six. Every other four-uniform class has $\mathcal C_3\ge12$, and $\delta\le2$ implies $\sum\delta\ge\mathcal C_3/2\ge6$. Thus Eq.~\eqref{eq:leakage_bound} is an exact minimum over four-uniform eleven-qutrit stabilizer resources with this reference-averaging convention. It is not a bound on arbitrary encodings or a claim of optimal recovery probability for a fixed reference.

For the attaining matrix, the six bad five-subsets all contain parties 1 and 2. Each other party belongs to two. Choosing either party 1 or 2 as reference leaves six uncorrectable four-erasure patterns out of 210; choosing any other party leaves two. The reference-averaged exact recovery fraction is therefore $76/77$, while each fixed-reference encoding corrects every erasure of at most three parties. The distribution of bad subsets, not only the scalar cost, determines these recovery statistics.

\section{Supporting constructions}
The small cases $\mathcal C_6^*(4)=2$ in the product family and $\mathcal C_2^*(7)=3$ recover established entanglement restrictions; Appendix 6 gives the witnesses and proofs \cite{higuchi2000entangled,gour2010all,huber2017absolutely,zhang2025extremal}. Unrestricted $\operatorname{AME}(4,6)$ states exist \cite{rather2022thirty,gross2025thirty}.

AfAME also produced a zero-cost $N=17$ certificate over $\mathbb F_{73}$ and $\mathbb F_{137}$, yielding local dimension $10001$. All 65,535 scored subsets have full rank in both sectors. Appendix 3 gives the matrices. Large-field AME existence is established by code-based constructions \cite{raissi2018optimal,feng2017multipartite,grassl2015quantum}; this is a reproducibility example, not a new existence regime.

\section{Search, verification, and scope}
AfAME uses symmetric edge updates, guided Metropolis--Hastings proposals, replica exchange, and restarts \cite{afame2026software}. For $(8,6)$ and $(11,3)$ the reported candidates were selected from seeds 42--49 and 42--69, respectively, with four replicas, 2000 proposals per replica per restart, and five restarts. An independent Python implementation recomputes every exported cut rank. These original search outputs are retained as baselines. The improved witnesses come from the published classifications and explicit sector alignments, rather than new heuristic discoveries. Independent checks establish their costs; the proofs in Secs.~\ref{sec:exact8} and \ref{sec:eleven} establish the respective optima. Appendix 2 specifies the search protocol. Researchers can supply their own matrices for verification without running a search, or export per-cut entropies and machine-readable certificates from a search. A separate Python implementation checks the certificates using explicit field arithmetic. This separation makes AfAME useful for reproducible construction and diagnosis within the supported families; it does not make a nonzero search result an optimality or nonexistence proof.

For composite searches, Eq.~\eqref{eq:min_separates} also implies a simple output improvement: retain the best matrix visited in each sector and recombine them, even if they were visited at different steps. The resulting cost cannot exceed the best jointly visited cost. Appendix 7 gives the guarantee and a held-out, matched-trajectory comparison over 150 runs. Its gains concern the export rule at a fixed proposal budget; they do not demonstrate faster construction or superiority to independently searching each field sector.

The exact calculations apply to the stated families and objective. For prime local dimensions, local Clifford equivalence transfers the entropy-deficit minima to stabilizer states. The composite result concerns products of the designated field sectors; no claim about arbitrary six-level states follows. The objective in Eq.~\eqref{eq:composite_cost} is sector resolved, rather than the square of the total entropy shortfall. Even a proven minimum leaves room for different entropy profiles, as Table~\ref{tab:ternary_profiles} demonstrates.

The cut-rank relation is established prior work. Our finite classification evaluations inherit their coverage from published inputs, with independent arithmetic checks and supplied data enabling reproduction. Each cut rank is polynomial to compute, but the number of subsets is exponential in $N$. AfAME's search remains heuristic, and the examples do not establish a runtime advantage over other optimizers or code-based constructions. The eleven-qutrit argument shows how a uniformity threshold can reduce a global optimization to an optimal-distance classification; the remaining classification input is still essential.

\section{Conclusion}
The same optimal aggregate score can conceal different entanglement yields and different patterns of erasure leakage. For eight qutrits, the three squared-cost minimizing profiles form the complete rank-tail Pareto boundary, giving sharp probabilities for exact blockwise entanglement extraction. The connection to weight enumerators makes the overlap with average-purity optimization explicit. For eleven qutrits, uniformity propagation and a published optimal-distance classification establish the exact minimum six and a sharp reference-averaged leakage bound. In local dimension six, sector alignment separates scalar optimality from maximizing fully mixed reductions, with exact stabilizer-product boundary $(44,48)$ and $(48,56)$ and a count maximum 56 valid for arbitrary pure binary--ternary products.

These are consequences of established stabilizer structure and classifications, supported by explicit certificates and independent arithmetic. They also show how AfAME remains useful when AME construction is unavailable: it can return objective-specific MMES witnesses and certify their rank profiles, costs, and operational counts within the stated family. They do not decide unrestricted AME existence or select a universally preferred entanglement objective. They instead determine precisely which performance trade-offs can be achieved within the stated families.

\section*{Acknowledgments}
Z.D. acknowledges funding from UK Research and Innovation under the UK government's Horizon Europe funding guarantee (grant EP/Z000807/1). 
OpenAI Codex (GPT-6 Astra) was used as auxiliary tool to assist with code refactoring, verification scripts, and literature mapping. The author conceived and led the research and verified the results.

\section*{Software}
The AfAME code and an independent Python verifier are available at  \href{https://doi.org/10.5281/zenodo.23068891}{DOI: 10.5281/zenodo.23068891} \cite{afame2026software}. 

\section*{Data availability}
The ternary input records are from Danielsen's public database \cite{danielsenDatabase}. The data that support the findings of this article are openly available in the \texttt{data/} directory of AfAME.

\section*{Conflicts of interest}
The author declares no conflicts of interest.

\bibliographystyle{apsrev4-2}
\bibliography{refs}

\onecolumngrid
\section*{Appendices}

\setcounter{subsection}{0}
\section*{Appendix 1: Character-sum derivation of the cut-rank identity}
\label{app:rpd-proof}
Consider the finite-field phase state defined in the main text: 
\begin{equation}
\ket{\Phi_P} = \lvert\mathbb{F}\rvert^{-N/2} \sum_{\mathbf{q}\in\mathbb{F}^N} \chi\bigl(\phi_P(\mathbf{q})\bigr) \ket{\mathbf{q}} .
\end{equation}
Split the system into two subsets; $S$ ($|S| =k$) and $\bar S$ ($|\bar S|=\ell=N-k$), and write $\mathbf{q} = (\mathbf{q}_S, \mathbf{q}_{\bar S})$ with $\mathbf{q}_S\in\mathbb{F}^k$, $\mathbf{q}_{\bar S}\in\mathbb{F}^{\ell}$.
Writing $\Delta\phi=\phi_P(\mathbf q_S,\mathbf q_{\bar S})-\phi_P(\mathbf q'_S,\mathbf q_{\bar S})$, the reduced density matrix is
\begin{equation}
\rho_S = \operatorname{Tr}_{\bar S}\bigl(\ket{\Phi_P}\!\bra{\Phi_P}\bigr) 
= \lvert\mathbb{F}\rvert^{-N} \sum_{\mathbf{q}_S,\mathbf{q}_S'} \sum_{\mathbf{q}_{\bar S}}
   \chi(\Delta\phi)
   \ket{\mathbf{q}_S}\!\bra{\mathbf{q}_S'} .
\end{equation}

\noindent \textbf{Splitting the quadratic form.} We decompose $\phi_P(\mathbf{q})$ into terms that involve only $S$, only $\bar S$, and their mixture:
\begin{align}
\phi_P(\mathbf{q}) &= \underbrace{\sum_{i\in S} P_{ii} q_i^2 + \sum_{i<j\in S} P_{ij} q_i q_j}_{\displaystyle \phi_S(\mathbf{q}_S)} \\
&\quad + \underbrace{\sum_{i\in\bar S} P_{ii} q_i^2 + \sum_{i<j\in\bar S} P_{ij} q_i q_j}_{\displaystyle \phi_{\bar S}(\mathbf{q}_{\bar S})} \nonumber\\
&\quad + \sum_{i\in S,\, j\in\bar S} P_{ij} q_i q_j .
\end{align}
The mixed part is exactly $\mathbf{q}_S^{\mathsf T} P_{S,\bar{S}}\,\mathbf{q}_{\bar S}$, where $P_{S,\bar{S}}$ is the $k\times\ell$ cut of $P$ formed by rows in $S$ and columns in $\bar S$ (the diagonal of $P$ does not enter here, only the off-diagonal block).
Thus
\begin{equation}
\phi_P(\mathbf{q}) = \phi_S(\mathbf{q}_S) + \phi_{\bar S}(\mathbf{q}_{\bar S}) + \mathbf{q}_S^{\mathsf T} P_{S,\bar{S}}\,\mathbf{q}_{\bar S}.
\end{equation}

The difference appearing in the exponent becomes
\begin{multline}
\Delta\phi = \phi_P(\mathbf{q}_S,\mathbf{q}_{\bar S}) - \phi_P(\mathbf{q}_S',\mathbf{q}_{\bar S}) \\
= \bigl[\phi_S(\mathbf{q}_S)-\phi_S(\mathbf{q}_S')\bigr]
   + (\mathbf{q}_S-\mathbf{q}_S')^{\mathsf T} P_{S,\bar{S}}\,\mathbf{q}_{\bar S}.
\end{multline}
Crucially, the $\phi_{\bar S}$ part cancels entirely, and the diagonal entries of $P$ appear only inside $\phi_S$ or $\phi_{\bar S}$, never in the cross term.
Define $\Delta\mathbf{q}_S = \mathbf{q}_S-\mathbf{q}_S'$ and $\mathbf{a} = P_{S,\bar{S}}^{\mathsf T}\,\Delta\mathbf{q}_S \in \mathbb{F}^{\ell}$.
Then $(\Delta\mathbf{q}_S)^{\mathsf T} P_{S,\bar{S}}\,\mathbf{q}_{\bar S} = \mathbf{a}\cdot\mathbf{q}_{\bar S}$, where the dot product is the natural pairing over $\mathbb{F}$.

\noindent \textbf{Summation over $\mathbf{q}_{\bar S}$: character orthogonality \cite{lidl1997finite}.}
Because $\chi$ is a homomorphism, the sum over $\mathbf{q}_{\bar S}$ factorises:
\begin{equation}
\sum_{\mathbf{q}_{\bar S}\in\mathbb{F}^{\ell}} \chi\bigl(\mathbf{a}\cdot\mathbf{q}_{\bar S}\bigr)
= \prod_{j\in\bar S}\Bigl( \sum_{x\in\mathbb{F}} \chi(a_j x) \Bigr).
\end{equation}
For any $a\in\mathbb{F}$, the additive character satisfies
\begin{equation}
\sum_{x\in\mathbb{F}} \chi(a x) =
\begin{cases}
\lvert\mathbb{F}\rvert & \text{if } a = 0,\\
0 & \text{if } a \neq 0 .
\end{cases}
\end{equation}
Hence the product is non-zero \textbf{only if} every $a_j = 0$, i.e.\ $\mathbf{a} = \mathbf{0}$; in that case it equals $\lvert\mathbb{F}\rvert^{\ell}$.
The condition $\mathbf{a}=\mathbf{0}$ is equivalent to
\begin{equation}
P_{S,\bar{S}}^{\mathsf T}\,\Delta\mathbf{q}_S = \mathbf{0}
\quad\Longleftrightarrow\quad
\Delta\mathbf{q}_S \in \ker\bigl(P_{S,\bar{S}}^{\mathsf T}\bigr).
\end{equation}

\noindent \textbf{Surviving matrix elements.}
When the kernel condition holds, the sum over $\mathbf{q}_{\bar S}$ gives $\lvert\mathbb{F}\rvert^{\ell}$, and the remaining factor $\chi(\phi_S(\mathbf{q}_S)-\phi_S(\mathbf{q}_S'))$ is a pure phase of modulus $1$.  Therefore
\begin{equation}
\bra{\mathbf{q}_S}\rho_S\ket{\mathbf{q}_S'}
= \frac{\mathbf{1}_{\{P_{S,\bar S}^{\mathsf T}(\mathbf{q}_S-\mathbf{q}_S')=0\}}}{\lvert\mathbb{F}\rvert^{k}}\, \chi(\phi_S(\mathbf{q}_S)-\phi_S(\mathbf{q}_S')).
\end{equation}

\noindent \textbf{Reduced-state spectrum.}
Let $K=\ker(P_{S,\bar S}^{\mathsf T})$ and let $D$ be diagonal with entries $\chi(\phi_S(\mathbf q_S))$. Then $D^\dagger\rho_S D$ removes the phase in the preceding equation. The resulting matrix has a block for each coset of $K$ in $\mathbb F^k$. Every block is the $|K|\times|K|$ all-ones matrix divided by $q^k$, with one nonzero eigenvalue $|K|/q^k=q^{-r_S}$. There are $q^k/|K|=q^{r_S}$ cosets. This proves the spectrum and both entropy formulas stated in the main text without invoking stabilizer generators.

\noindent \textbf{Purity calculation.}
The purity of $\rho_S$ is
\begin{equation}
\operatorname{Tr}(\rho_S^2) = \sum_{\mathbf{q}_S,\mathbf{q}_S'} \bigl|\bra{\mathbf{q}_S}\rho_S\ket{\mathbf{q}_S'}\bigr|^2 .
\end{equation}
Only pairs satisfying the kernel condition contribute, and for those the absolute value removes the phase, giving $\lvert\mathbb{F}\rvert^{-2k}$.  Thus
\begin{equation}
\operatorname{Tr}(\rho_S^2) = \lvert\mathbb{F}\rvert^{-2k}\; \times\; \#\bigl\{ (\mathbf{q}_S,\mathbf{q}_S') : \Delta\mathbf{q}_S \in \ker(P_{S,\bar{S}}^{\mathsf T}) \bigr\}. 
\label{eq:purity}
\end{equation}

\noindent \textbf{Counting the valid pairs.}
The set $\ker(P_{S,\bar{S}}^{\mathsf T})$ is a linear subspace of $\mathbb{F}^{k}$.
By the rank-nullity theorem over $\mathbb{F}$,
\begin{equation}
\dim_{\mathbb{F}} \ker(P_{S,\bar{S}}^{\mathsf T}) = k - \operatorname{rk}_{\mathbb{F}}(P_{S,\bar{S}}^{\mathsf T}) = k - \operatorname{rk}_{\mathbb{F}}(P_{S,\bar{S}}).
\end{equation}
Hence $\lvert\ker(P_{S,\bar{S}}^{\mathsf T})\rvert = \lvert\mathbb{F}\rvert^{\,k - \operatorname{rk}(P_{S,\bar{S}})}$.
For each $\Delta\mathbf{q}_S \in \ker(P_{S,\bar{S}}^{\mathsf T})$, we may choose $\mathbf{q}_S$ arbitrarily ($\lvert\mathbb{F}\rvert^{k}$ choices) and set $\mathbf{q}_S' = \mathbf{q}_S - \Delta\mathbf{q}_S$.
Therefore the total number of ordered pairs is
\begin{equation}
\#\text{pairs} = \lvert\mathbb{F}\rvert^{k} \cdot \lvert\mathbb{F}\rvert^{\,k - \operatorname{rk}} = \lvert\mathbb{F}\rvert^{\,2k - \operatorname{rk}_{\mathbb{F}}(P_{S,\bar{S}})}.
\label{eq:counting}
\end{equation}

\noindent \textbf{Final purity formula.}
Inserting \eqref{eq:counting} into \eqref{eq:purity} yields
\begin{equation}
\operatorname{Tr}(\rho_S^2) = \lvert\mathbb{F}\rvert^{-2k} \cdot \lvert\mathbb{F}\rvert^{\,2k - \operatorname{rk}(P_{S,\bar{S}})}
= \lvert\mathbb{F}\rvert^{-\, \operatorname{rk}_{\mathbb{F}}(P_{S,\bar{S}})}.
\end{equation}
This yields the finite-field cut-rank purity identity.

\clearpage

\setcounter{subsection}{0}
\section*{Appendix 2: CRT structure and computational protocol}
The following details specify the composite-dimension phase family, its relation to ring-valued quadratic phases, and the exact verification and benchmark procedures.

\subsection{Composite local dimensions}
\subsubsection{Square-free dimensions: CRT factorisation}
The structure of $\mathbb{Z}_d$ depends on the arithmetic decomposition of the local dimension $d$.  In the square-free case, $d = \prod_{\alpha=1}^{r} p_\alpha$ with distinct primes $p_\alpha$, the Chinese Remainder Theorem (CRT) gives a ring isomorphism $\mathbb{Z}_d \cong \prod_{\alpha} \mathbb{F}_{p_\alpha}$. This induces a virtual tensor product decomposition on the single qudit Hilbert space. Consequently, the computational basis states $\ket{x}$ ($x\in\{0,\dots,d-1\}$) admit a tensor product representation in terms of lower-dimensional qudits.
\begin{align}
\mathbb{C}^d &\cong \bigotimes_{\alpha=1}^{r}\mathbb{C}^{p_\alpha},\\
U_{\rm CRT}\ket{x} &= \bigotimes_{\alpha=1}^{r}\ket{x_\alpha},
\qquad x_\alpha\equiv x\pmod{p_\alpha}.
\end{align}
The tensor product on the right-hand side is not a partition of the physical system into new parties. It is merely a reshuffling of the internal degrees of freedom of the same qudit. For $N$-physical parties, the global Hilbert space is the physical tensor product $\mathcal{H}_{\rm phys}
= (\mathbb{C}^d)^{\otimes N}$. By inserting the CRT decomposition of each qudit, the Hilbert space writes $\mathcal{H}_{\rm phys}
\;=\; \bigotimes_{\alpha=1}^{r} \Bigl( \bigotimes_{i=1}^{N} \mathbb{C}^{p_\alpha} \Bigr) \;\equiv\; \bigotimes_{\alpha=1}^{r} \mathcal{H}_{\alpha}$, where $\mathcal{H}_{\alpha} = (\mathbb{C}^{p_\alpha})^{\otimes N}$ is the Hilbert space of $N$ qudits of dimension $p_\alpha$. The $N$-party Hilbert space then factors into a product of $r$ independent prime sector Hilbert spaces, each describing the same $N$ parties but with a smaller local dimension. For a quadratic-phase state over $\mathbb{Z}_d$ with the standard character, the local CRT basis map gives an exact product of sector states. Set $a_\alpha=(d/p_\alpha)^{-1}\pmod{p_\alpha}$. Then
\begin{equation}
e^{2\pi i x/d}=\prod_\alpha e^{2\pi i a_\alpha x_\alpha/p_\alpha},
\qquad P^{(\alpha)}=a_\alpha(P\bmod p_\alpha).
\end{equation}
Every $a_\alpha$ is nonzero, so this scaling preserves cut ranks and subsystem entropies. The unscaled residue matrix need not give the same sector state under this basis map.
\begin{equation}
  U_{\rm CRT}^{\otimes N}\ket{\Phi_P}
  =\bigotimes_{\alpha=1}^{r}\ket{\Phi_{P^{(\alpha)}}}_{p_\alpha}.
  \label{eq:crt_state}
\end{equation}

\begin{figure*}[t]
\centering
\resizebox{0.65\textwidth}{!}
{
\begin{tikzpicture}[
  font=\small,
  >=Stealth,
  qudit/.style={
    draw, rounded corners=1.5pt, line width=0.4pt,
    minimum width=0.9cm, minimum height=0.60cm,
    align=center, font=\scriptsize, inner sep=1.2pt
  },
  ca/.style={fill=blue!10,   draw=blue!65,   line width=0.5pt},
  cb/.style={fill=red!10,    draw=red!60,    line width=0.5pt},
  cc/.style={fill=orange!10, draw=orange!65, line width=0.5pt},
  cd/.style={fill=teal!12,   draw=teal!65,   line width=0.5pt},
  braceU/.style={decorate,
    decoration={brace, amplitude=4pt, raise=2.5pt, mirror}},
  braceD/.style={decorate,
    decoration={brace, amplitude=4pt, raise=2.5pt, mirror}},
  lbl/.style={font=\scriptsize},
  plbl/.style={font=\normalsize\bfseries},
]

\begin{scope}[yshift=0.5cm]

\def\xA{0.0}   \def\xB{1.5}   \def\xC{3.0}   \def\xDot{4.2}   \def\xD{5.4}

\def\yI{0.0}   \def\yII{-1.0}   \def\yDot{-1.72}   \def\yN{-2.4}

\foreach \x/\col in {\xA/blue, \xB/red, \xC/orange, \xD/teal}
  \fill[\col!6] (\x-0.5, 0.40) rectangle (\x+0.5, \yN-0.38);

\node[lbl, black,   font=\scriptsize\bfseries] at (\xA,   0.62) {$\alpha=1$};
\node[lbl, black,    font=\scriptsize\bfseries] at (\xB,   0.62) {$\alpha=2$};
\node[lbl, black, font=\scriptsize\bfseries] at (\xC,   0.62) {$\alpha=3$};
\node[lbl]                                       at (\xDot, 0.62) {$\cdots$};
\node[lbl, black,   font=\scriptsize\bfseries] at (\xD,   0.62) {$\alpha=r$};

\foreach \y/\p in {0.0/1, -1.0/2, -2.4/N}
  \node[lbl, anchor=east] at (-0.9, \y) { $\p$};

\foreach \y/\p in {0.0/1, -1.0/2, -2.4/N}{
  \node[qudit, ca] at (\xA,   \y) {$\ket{x_1^{(\p)}}$};
  \node[qudit, cb] at (\xB,   \y) {$\ket{x_2^{(\p)}}$};
  \node[qudit, cc] at (\xC,   \y) {$\ket{x_3^{(\p)}}$};
  \node[lbl]       at (\xDot, \y) {$\cdots$};
  \node[qudit, cd] at (\xD,   \y) {$\ket{x_r^{(\p)}}$};
}

\foreach \x in {\xA, \xB, \xC, \xD}
  \node[font=\normalsize] at (\x, \yDot) {$\vdots$};

\foreach \y/\p in {0.0/1, -1.0/2, -2.4/N}
  \draw[braceU]
    (6.3, \y+0.3) -- (6.3, \y-0.3)
    node[midway, right=8pt, lbl] {$\ket{x^{(\p)}}_{d}$};

\def\ybb{-3.2}   

\draw[braceD]
  (\xA+0.50, \ybb) -- (\xA-0.50, \ybb)
  node[midway, below=7pt, lbl, black]
    {$\bigl(\mathbb{C}^{p_1}\bigr)^{\otimes N}$};

\draw[braceD]
  (\xB+0.50, \ybb) -- (\xB-0.50, \ybb)
  node[midway, below=7pt, lbl, black]
    {$\bigl(\mathbb{C}^{p_2}\bigr)^{\otimes N}$};

\draw[braceD]
  (\xC+0.50, \ybb) -- (\xC-0.50, \ybb)
  node[midway, below=7pt, lbl, black]
    {$\bigl(\mathbb{C}^{p_3}\bigr)^{\otimes N}$};

\draw[braceD]
  (\xD+0.50, \ybb) -- (\xD-0.50, \ybb)
  node[midway, below=7pt, lbl, black]
    {$\bigl(\mathbb{C}^{p_r}\bigr)^{\otimes N}$};

\end{scope}
\end{tikzpicture}
}
\caption{\textbf{CRT structure for square-free $d$.}
$N$ parties $\times$ $r$ prime factors grid. \textit{Horizontal}: CRT recombines the virtual prime-sector qudits of a single party into one $d$-level qudit. \textit{Vertical}: each column is an independent $N$-party state on a prime-sector Hilbert space.}
\label{fig:sectorProd}
\end{figure*}

\subsubsection{Purity factorisation and entropy additivity}

From Eq.~\eqref{eq:crt_state}, the reduced density matrix on any subsystem $S$ factorizes \emph{after the partywise CRT basis change}:
\begin{align}
U_{\rm CRT}^{\otimes |S|}\rho_S
(U_{\rm CRT}^{\dagger})^{\otimes |S|}
&= \bigotimes_{\alpha=1}^{r} \rho_S^{(\alpha)},\\
\rho_S^{(\alpha)}
&= \operatorname{Tr}_{\bar{S}}\bigl(
\ket{\Phi_{P^{(\alpha)}}}\!\bra{\Phi_{P^{(\alpha)}}}_{p_\alpha}
\bigr).
\end{align}
The local basis change preserves the spectrum, so the purity factorizes:
\begin{equation}
  \operatorname{Tr}(\rho_S^2) = \prod_{\alpha=1}^{r}
    \operatorname{Tr}\bigl((\rho_S^{(\alpha)})^2\bigr)
  = \prod_{\alpha=1}^{r} p_\alpha^{-\operatorname{rk}_{\mathbb{F}_{p_\alpha}}\!(P^{(\alpha)}_{S,\bar{S}})}.
  \label{eq:composite_purity}
\end{equation}
The AME condition for the composite state is that each factor satisfies the saturation condition:
\begin{equation}
  \operatorname{rk}_{\mathbb{F}_{p_\alpha}}\!(P^{(\alpha)}_{S,\bar{S}}) = |S|
  \quad \forall\,\alpha,\ \forall S\ \text{with}\ |S|\le N/2.
  \label{eq:composite_ame}
\end{equation}
The flat spectra in Eq.~\eqref{eq:rpd} imply that von Neumann and second R\'enyi entropies coincide here. Both become sums over prime sectors:
\begin{equation}
  S(\rho_S)=S_2(\rho_S)
  = \sum_{\alpha=1}^{r} \operatorname{rk}_{\mathbb{F}_{p_\alpha}}\!(P^{(\alpha)}_{S,\bar{S}})\; \ln p_\alpha,
  \label{eq:entropy_sum}
\end{equation}
which for an AME state evaluates to $|S|\sum_\alpha\ln p_\alpha = |S|\ln d$, the maximal possible value. Within this product family, construction at square-free $d$ decouples into independent prime-field problems.

Entropy additivity in Eq.~\eqref{eq:entropy_sum} is the standard tensor-product identity, specialized here to the known flat cut-rank spectra. The product reduction is maximally mixed if and only if every prime-field factor is maximally mixed on the same cut. This is a property of the CRT-product family, not an obstruction for arbitrary $d$-level states.

\subsubsection{General mixed dimensions}
The same character sum over the ring $\mathbb{Z}_d$ gives a kernel-counting formula, rather than a field-rank formula:
\begin{equation}
\operatorname{Tr}(\rho_S^2)=\frac{|\ker(P_{S,\bar S}^{\mathsf T}:\mathbb{Z}_d^{|S|}\to\mathbb{Z}_d^{N-|S|})|}{d^{|S|}}.
\end{equation}
For a prime-power ring $\mathbb{Z}_{p^m}$, the cut map is injective if and only if it has a $k\times k$ minor whose determinant is not divisible by $p$, where $k=|S|\le N-|S|$. This is equivalent to full row rank of the reduced matrix over $\mathbb{F}_p$. CRT then gives the following AME criterion for ring-valued quadratic-phase states:

\begin{equation}
\operatorname{rk}_{\mathbb{F}_{p_i}}\!\bigl(P_{S,\bar{S}}\bmod p_i\bigr) = |S|
\qquad \forall\, i,\ \forall S\ \text{with}\ |S|\le N/2.
\label{eq:mixed}
\end{equation}

This gives an AME injectivity criterion for ring-valued quadratic-phase states in composite local dimensions 
($d = \prod_i p_i^{m_i}$), for instance, $d=12=3\times 2^2$. It should 
be noted, however, that the CRT factorisation holds at the level of 
Hilbert space tensor products and additive characters, while the 
rank--purity formula is strictly field-based unless additional structure 
is imposed; in the prime-power case $\mathbb{Z}_{p^m}$, the relevant 
condition is injectivity of the map $P_{S,\bar{S}}^{\mathsf{T}}$ over 
$\mathbb{Z}_{p^m}$, which reduces to full-rank of its mod-$p$ reduction 
over $\mathbb{F}_p$. For $m>1$, the ring $\mathbb Z_{p^m}$ is not the field $\mathbb F_{p^m}$. AfAME searches matrices over the latter and combines distinct field sectors as tensor factors; Eq.~\eqref{eq:mixed} concerns a separately defined ring-valued phase family and is not the program's prime-power field rank test.

\subsection{Search and certificate verification}
AfAME searches symmetric phase matrices by local edge updates, guided Metropolis--Hastings moves, replica exchange between temperature levels, and restarts \cite{afame2026software}. For the $(4,6)$ and $(7,2)$ examples we used four replicas, 2000 proposals per replica per restart, five restarts, and seed 42. For $(8,6)$ we used the same budget at seeds 42--49; for $(11,3)$ we used seeds 42--69. Each setting therefore used at most 40,000 proposals per seed. We selected a lowest-cost certificate and recomputed all required field cut ranks with a separate Python verifier that does not call the C++ search. The exact ranks determine the subsystem entropies through Eq.~\eqref{eq:rpd}; no sampling of quantum states or floating-point entropy estimation is used. The $(17,10001)$ certificate was checked on all 65,535 nonempty subsets of sizes at most eight in both prime sectors. Runtime comparisons and algorithmic speedup claims are outside the scope of these examples.

For the binary eight-party optimum, we generated all $1044$ nonisomorphic seven-vertex graphs from the NetworkX graph atlas \cite{read1998atlas}, then joined an eighth vertex to every subset of the first seven vertices. The diagonal of a phase matrix does not affect cut ranks, so these simple graphs cover the entire symmetric binary phase-matrix cost space up to vertex permutation. One program computes exact ranks by Gaussian elimination over $\mathbb F_2$; a separately written program computes them from the cardinality of the image of every row combination. Both cover all $133{,}632$ extensions and obtain minimum cost $28$; the elimination program permits safe partial-cost pruning, while the row-span program evaluates every cut. The input atlas rows, both source programs, their output, and an independently verified witness are included in accompanying data. Duplicated eight-vertex isomorphism classes do not affect the minimum.

\subsection{Recombination benchmark and reproducibility}
The benchmark protocol was fixed after a separate exploratory set of seeds 200--219 and before the held-out runs. Held-out runs used $d=6$, $N\in\{10,11,12\}$, seeds 1000--1049, four replicas, 500 steps per replica, one restart, temperature endpoints 0.25 and 50, guidance parameter 0.85, exchange interval 10, and logging interval 500. The instrumented C++ version stores sector records after initialization, accepted updates, exchanges, and each restart; recording and recombination use no new random draws. Nine baseline reruns, three seeds at each $N$, matched the instrumented joint best costs exactly. A separate Python program recomputed every exported cost from its field matrices and all required cuts. The 95\% intervals in Appendix 7 are percentile intervals from 50,000 bootstrap resamples of the 50 seedwise paired differences at each $N$, using NumPy random seed 20260923. No test of runtime or cross-algorithm superiority was prespecified. The frozen protocol, 150 certificates, per-seed outcomes, instrumented code patch, and plotting data accompany this manuscript.

\setcounter{subsection}{0}
\section*{Appendix 3: Phase matrix decomposition: \texorpdfstring{$N=17$, $d=10001$}{N=17, d=10001}} \label{app:ame17}
We explicitly provide the phase matrices for the AME$(17, 10001)$ state as follows: By the Chinese Remainder Theorem, the global phase matrix over $\mathbb{Z}_{10001}$ has the following residue matrices over $\mathbb{F}_{73}$ and $\mathbb{F}_{137}$. For the standard additive characters, the actual tensor-factor phase matrices include the nonzero character-weight scalars described in Appendix 2; these scalars leave every cut rank unchanged.

\textbf{Factor 1:} $P_{\mathbb{F}_{73}}$

\begingroup\small\setlength{\arraycolsep}{2.6pt}
\begin{equation*}
\left(\begin{array}{*{17}{r}}
0 & 14 & 59 & 54 & 37 & 6 & 8 & 34 & 17 & 65 & 21 & 44 & 51 & 38 & 5 & 37 & 40 \\
14 & 0 & 52 & 43 & 20 & 26 & 54 & 43 & 39 & 54 & 37 & 6 & 17 & 28 & 38 & 43 & 49 \\
59 & 52 & 0 & 15 & 53 & 68 & 33 & 27 & 63 & 26 & 61 & 14 & 8 & 40 & 27 & 37 & 71 \\
54 & 43 & 15 & 0 & 68 & 60 & 15 & 42 & 41 & 47 & 27 & 8 & 55 & 66 & 12 & 59 & 35 \\
37 & 20 & 53 & 68 & 0 & 71 & 33 & 54 & 62 & 47 & 8 & 1 & 59 & 61 & 23 & 70 & 6 \\
6 & 26 & 68 & 60 & 71 & 0 & 37 & 59 & 69 & 72 & 34 & 7 & 48 & 71 & 52 & 31 & 60 \\
8 & 54 & 33 & 15 & 33 & 37 & 0 & 56 & 33 & 8 & 21 & 13 & 14 & 24 & 45 & 41 & 42 \\
34 & 43 & 27 & 42 & 54 & 59 & 56 & 0 & 56 & 50 & 55 & 36 & 56 & 10 & 42 & 43 & 44 \\
17 & 39 & 63 & 41 & 62 & 69 & 33 & 56 & 0 & 19 & 30 & 51 & 20 & 30 & 32 & 25 & 39 \\
65 & 54 & 26 & 47 & 47 & 72 & 8 & 50 & 19 & 0 & 52 & 66 & 2 & 61 & 25 & 2 & 25 \\
21 & 37 & 61 & 27 & 8 & 34 & 21 & 55 & 30 & 52 & 0 & 21 & 11 & 61 & 57 & 38 & 0 \\
44 & 6 & 14 & 8 & 1 & 7 & 13 & 36 & 51 & 66 & 21 & 0 & 58 & 63 & 47 & 55 & 61 \\
51 & 17 & 8 & 55 & 59 & 48 & 14 & 56 & 20 & 2 & 11 & 58 & 0 & 67 & 56 & 6 & 71 \\
38 & 28 & 40 & 66 & 61 & 71 & 24 & 10 & 30 & 61 & 61 & 63 & 67 & 0 & 42 & 32 & 14 \\
5 & 38 & 27 & 12 & 23 & 52 & 45 & 42 & 32 & 25 & 57 & 47 & 56 & 42 & 0 & 57 & 35 \\
37 & 43 & 37 & 59 & 70 & 31 & 41 & 43 & 25 & 2 & 38 & 55 & 6 & 32 & 57 & 0 & 49 \\
40 & 49 & 71 & 35 & 6 & 60 & 42 & 44 & 39 & 25 & 0 & 61 & 71 & 14 & 35 & 49 & 0
\end{array}\right)
\end{equation*}
\endgroup

\textbf{Factor 2:} $P_{\mathbb{F}_{137}}$
\begingroup\small\setlength{\arraycolsep}{2.6pt}
\begin{equation*}
\left(\begin{array}{*{17}{r}}
0 & 87 & 48 & 100 & 28 & 35 & 90 & 1 & 36 & 51 & 100 & 7 & 127 & 69 & 6 & 122 & 76 \\
87 & 0 & 42 & 103 & 120 & 68 & 12 & 117 & 124 & 102 & 6 & 73 & 18 & 80 & 17 & 81 & 19 \\
48 & 42 & 0 & 93 & 8 & 84 & 4 & 43 & 130 & 45 & 116 & 81 & 10 & 106 & 72 & 23 & 65 \\
100 & 103 & 93 & 0 & 125 & 112 & 20 & 100 & 1 & 87 & 89 & 66 & 65 & 128 & 102 & 52 & 1 \\
28 & 120 & 8 & 125 & 0 & 125 & 44 & 63 & 74 & 25 & 123 & 47 & 109 & 78 & 51 & 13 & 43 \\
35 & 68 & 84 & 112 & 125 & 0 & 44 & 25 & 79 & 31 & 67 & 56 & 100 & 133 & 109 & 124 & 88 \\
90 & 12 & 4 & 20 & 44 & 44 & 0 & 12 & 107 & 134 & 136 & 134 & 76 & 7 & 37 & 17 & 81 \\
1 & 117 & 43 & 100 & 63 & 25 & 12 & 0 & 54 & 7 & 98 & 57 & 133 & 135 & 77 & 4 & 2 \\
36 & 124 & 130 & 1 & 74 & 79 & 107 & 54 & 0 & 127 & 10 & 30 & 133 & 16 & 57 & 43 & 129 \\
51 & 102 & 45 & 87 & 25 & 31 & 134 & 7 & 127 & 0 & 92 & 48 & 87 & 132 & 15 & 101 & 27 \\
100 & 6 & 116 & 89 & 123 & 67 & 136 & 98 & 10 & 92 & 0 & 87 & 109 & 76 & 53 & 9 & 36 \\
7 & 73 & 81 & 66 & 47 & 56 & 134 & 57 & 30 & 48 & 87 & 0 & 12 & 39 & 74 & 86 & 67 \\
127 & 18 & 10 & 65 & 109 & 100 & 76 & 133 & 133 & 87 & 109 & 12 & 0 & 120 & 46 & 128 & 23 \\
69 & 80 & 106 & 128 & 78 & 133 & 7 & 135 & 16 & 132 & 76 & 39 & 120 & 0 & 86 & 38 & 67 \\
6 & 17 & 72 & 102 & 51 & 109 & 37 & 77 & 57 & 15 & 53 & 74 & 46 & 86 & 0 & 86 & 4 \\
122 & 81 & 23 & 52 & 13 & 124 & 17 & 4 & 43 & 101 & 9 & 86 & 128 & 38 & 86 & 0 & 102 \\
76 & 19 & 65 & 1 & 43 & 88 & 81 & 2 & 129 & 27 & 36 & 67 & 23 & 67 & 4 & 102 & 0
\end{array}\right)
\end{equation*}
\endgroup

\vspace{1em}

\noindent \textbf{CRT collapsed system:} $P_{\mathbb{Z}_{10001}}$
\begingroup\small\setlength{\arraycolsep}{2.6pt}
\begin{equation*}
\left(\begin{array}{*{17}{r}}
0 & 87 & 2103 & 9690 & 9892 & 8255 & 227 & 6166 & 9215 & 5394 & 3525 & 555 & 6840 & 6097 & 8911 & 6972 & 624 \\
87 & 0 & 1001 & 4350 & 531 & 4041 & 6040 & 9022 & 6974 & 7500 & 3979 & 6649 & 8923 & 3094 & 3031 & 8438 & 2896 \\
2103 & 1001 & 0 & 4614 & 9324 & 2550 & 1785 & 2509 & 6706 & 9224 & 9843 & 6657 & 7819 & 7778 & 757 & 5366 & 6641 \\
9690 & 4350 & 4614 & 0 & 7660 & 3126 & 4541 & 6539 & 3837 & 6252 & 2144 & 6505 & 9107 & 2183 & 1472 & 7724 & 7262 \\
9892 & 531 & 9324 & 7660 & 0 & 947 & 7990 & 200 & 6924 & 4135 & 4096 & 9637 & 5315 & 1448 & 9367 & 2479 & 9496 \\
8255 & 4041 & 2550 & 3126 & 947 & 0 & 2373 & 7286 & 9121 & 4963 & 3903 & 6358 & 3114 & 2188 & 7644 & 8207 & 9404 \\
227 & 6040 & 1785 & 4541 & 7990 & 2373 & 0 & 8232 & 9012 & 2052 & 4109 & 7532 & 2131 & 8638 & 8805 & 6319 & 7342 \\
6166 & 9022 & 2509 & 6539 & 200 & 7286 & 8232 & 0 & 2246 & 7131 & 2975 & 7044 & 5750 & 3149 & 1721 & 2744 & 6030 \\
9215 & 6974 & 6706 & 3837 & 6924 & 9121 & 9012 & 2246 & 0 & 1771 & 1928 & 3044 & 6298 & 5359 & 2660 & 317 & 1499 \\
5394 & 7500 & 9224 & 6252 & 4135 & 4963 & 2052 & 7131 & 1771 & 0 & 6257 & 9775 & 6937 & 2324 & 974 & 1608 & 7836 \\
3525 & 3979 & 9843 & 2144 & 4096 & 3903 & 4109 & 2975 & 1928 & 6257 & 0 & 7759 & 2712 & 3638 & 4437 & 1790 & 584 \\
555 & 6649 & 6657 & 6505 & 9637 & 6358 & 7532 & 7044 & 3044 & 9775 & 7759 & 0 & 423 & 6341 & 485 & 6114 & 3492 \\
6840 & 8923 & 7819 & 9107 & 5315 & 3114 & 2131 & 5750 & 6298 & 6937 & 2712 & 423 & 0 & 2038 & 1005 & 6430 & 2626 \\
6097 & 3094 & 7778 & 2183 & 1448 & 2188 & 8638 & 3149 & 5359 & 2324 & 3638 & 6341 & 2038 & 0 & 1867 & 3463 & 1985 \\
8911 & 3031 & 757 & 1472 & 9367 & 7644 & 8805 & 1721 & 2660 & 974 & 4437 & 485 & 1005 & 1867 & 0 & 8306 & 3977 \\
6972 & 8438 & 5366 & 7724 & 2479 & 8207 & 6319 & 2744 & 317 & 1608 & 1790 & 6114 & 6430 & 3463 & 8306 & 0 & 2020 \\
624 & 2896 & 6641 & 7262 & 9496 & 9404 & 7342 & 6030 & 1499 & 7836 & 584 & 3492 & 2626 & 1985 & 3977 & 2020 & 0
\end{array}\right)
\end{equation*}
\endgroup

\begin{table*}[tb]
\centering
\caption{Certification of the $\text{AME}(17,10001)$ state ($d=10001$). 
For every subsystem size $|S|$ (denoted $k$ in the table), all $n_{\text{bips}}$ bipartitions achieve full-rank, so the ratio 
$\mathcal{R} = \operatorname{rk}(P_{S,\bar{S}})/|S| = 1$. 
This certifies that the state is AME within the stated quadratic-phase construction. Encoded logical-qudit code parameters require a separate construction.}

\begin{tabular}{c|c|c|c|c|c}
Size $k$ & Subsets & Target Rank & Rank Saturation & Ratio $\mathcal{R}$ & Status \\ \hline
1 & 17 & 1 & 17/17 & 1.000 & \checkmark \\
2 & 136 & 2 & 136/136 & 1.000 & \checkmark \\
3 & 680 & 3 & 680/680 & 1.000 & \checkmark \\
4 & 2,380 & 4 & 2,380/2,380 & 1.000 & \checkmark \\
5 & 6,188 & 5 & 6,188/6,188 & 1.000 & \checkmark \\
6 & 12,376 & 6 & 12,376/12,376 & 1.000 & \checkmark \\
7 & 19,448 & 7 & 19,448/19,448 & 1.000 & \checkmark \\
8 & 24,310 & 8 & 24,310/24,310 & 1.000 & \checkmark \\
\hline
\textbf{Total} & \textbf{65,535} & -- & \textbf{100\% Perfect} & \textbf{1.0} & \textbf{CERTIFIED} \\
\end{tabular}

\end{table*}

\setcounter{subsection}{0}
\section*{Appendix 4: Exact binary eight-party optimum}
For $N=8$ and $q=2$, the diagonal entries of the symmetric phase matrix never occur in a cut block. Hence every cost is represented by the adjacency matrix of a simple eight-vertex graph. To cover these graphs up to isomorphism, start from the complete atlas of $1044$ nonisomorphic seven-vertex graphs \cite{read1998atlas}. For each atlas graph, attach one new vertex with all $2^7=128$ possible neighbourhoods. Every eight-vertex graph is isomorphic to an extension in this set: delete any one vertex, identify the remaining seven-vertex graph with an atlas member, and restore the deleted vertex and its incident edges. Vertex relabelling preserves the objective.

We covered all $1044\times128=133{,}632$ extensions, including duplicates, for the objective containing $\sum_{k=1}^4\binom8k=162$ required subsets. Two independent implementations gave the same minimum, $\mathcal C_2^*(8)=28$: one uses Gaussian elimination over $\mathbb F_2$, and the other enumerates the images of all row combinations to obtain each cut rank. The lower bound follows from this exhaustive coverage; the following witness attains it. With vertices in row and column order $1,\ldots,8$, its adjacency matrix is
\begin{equation*}
\left(\begin{array}{cccccccc}
0&1&0&0&0&1&1&1\\
1&0&1&0&0&0&0&1\\
0&1&0&1&0&0&0&1\\
0&0&1&0&1&0&1&0\\
0&0&0&1&0&1&0&1\\
1&0&0&0&1&0&0&1\\
1&0&0&1&0&0&0&1\\
1&1&1&0&1&1&1&0
\end{array}\right).
\end{equation*}
The witness has 140 full-rank cuts, 20 cuts with deficit one, and two with deficit two, all deficiencies occurring at size four; its objective is $20\cdot1^2+2\cdot2^2=28$. AfAME's binary matrix in the $(8,6)$ certificate has the same deficit histogram and cost. The elimination program may stop a candidate evaluation once its nonnegative partial cost exceeds the current minimum; the independent span program evaluates every cut without pruning. Both re-executions returned minimum 28 and 375 attaining extensions (these extensions are not distinct equivalence classes). Cabello \emph{et al.} already tabulated bipartite Schmidt ranks for eight-qubit graph-state classes \cite{cabello2009eight}; this computation evaluates the particular squared-deficit objective on that established entanglement structure. Appendix 5 gives the ternary optimum 16, which together with this binary result establishes the product minimum 44.

\setcounter{subsection}{0}
\section*{Appendix 5: Exact ternary eight-party optimum}
\label{app:ternary8}
\subsection{Coverage and input provenance}
Danielsen's classification gives one representative of each length-eight self-dual additive code over $\mathbb F_9$ under coordinate permutations and independent symplectic transformations of the coordinates \cite{danielsen2009graph,danielsen2012classification}. For qutrit stabilizer states these are party permutations and local Clifford operations. Every class has a weighted graph representative, and the published database includes all 817 classes, including 158 decomposable classes. The cost is invariant under these operations, so evaluating one representative per class covers the entire optimization over ternary phase matrices. Our use of the published classification does not independently regenerate its equivalence classes.

We downloaded \texttt{selfdualcodes3.txt.bz2} from the database \cite{danielsenDatabase} and extracted every record with length eight. The source-archive and extracted-record SHA-256 checksums, original line numbers, and exact input records are supplied in the data package. The original compressed archive is not redistributed; its URL and checksum are recorded in the provenance file. The extraction checks the count 817. The evaluation also checks the indecomposable/decomposable split and the mass identity
\begin{equation}
\sum_{j=1}^{817}\frac{24^8\,8!}{|\operatorname{Aut}(C_j)|}
=\prod_{i=1}^{8}(3^i+1)
=234870301468364800.
\end{equation}
This uses the published automorphism orders as an input-integrity check; it is not a new computation of those orders or an independent proof of classification completeness.

\subsection{Independent calculations}
The Python program parses the weighted graphs and uses the separate AfAME certificate verifier's finite-field elimination. It evaluates every cut $1\le|S|\le4$ and writes the class index, subset mask, rank, and deficit. A separately implemented C++ program parses the source records and enumerates all $x\in\mathbb F_3^8$. The corresponding graph stabilizer label is $(x,Px)$, with support
\begin{equation}
\operatorname{supp}(x,Px)=\{i:(x_i,(Px)_i)\ne(0,0)\}.
\end{equation}
For an element to be supported inside $S$, $x_{\bar S}=0$ and $P_{\bar S,S}x_S=0$ are necessary and sufficient. Therefore its support count satisfies
\begin{equation}
\#\{x:\operatorname{supp}(x,Px)\subseteq S\}
=3^{|S|-\operatorname{rk}_{\mathbb F_3}P_{S,\bar S}}
=3^{\delta_S}.
\end{equation}
The second program obtains these counts by summing exact support multiplicities over submasks; it does not call an elimination routine. It also reproduces the published full weight enumerator and minimum distance of every one of the 817 classes. All 132,354 cut ranks agree between programs.

\begin{table}[tb]
\caption{Exhaustive ternary evaluation grouped by the stabilizer's minimum nonzero support weight $d_{\rm code}$. Both decomposable and indecomposable classes are included.}
\centering
\begin{tabular}{rrr}
\toprule
$d_{\rm code}$ & Classes & Minimum $\mathcal C_3$\\
\midrule
1 & 121 & 99\\
2 & 424 & 41\\
3 & 195 & 23\\
4 & 77 & 16\\
\bottomrule
\end{tabular}
\end{table}

\subsection{Witnesses and entropy profiles}
A minimizing adjacency matrix, with rows and columns labelled $1,\ldots,8$, is
\begin{equation}
P_\star^{(3)}=\begin{pmatrix}
0&2&0&1&0&0&2&2\\
2&0&1&0&1&0&2&2\\
0&1&0&0&0&1&2&2\\
1&0&0&0&0&1&2&1\\
0&1&0&0&0&2&2&1\\
0&0&1&1&2&0&2&1\\
2&2&2&2&2&2&0&0\\
2&2&2&1&1&1&0&0
\end{pmatrix}.
\end{equation}
It is record 741 in the extracted file (one-based indexing). All 92 smaller cuts have full rank; the balanced cuts have ranks four on 54 subsets and three on 16 subsets. The existing C++ AfAME verification-only mode independently confirms cost 16 for this matrix.

Records 741--750 attain the minimum. Seven have balanced deficit counts $(54,16,0)$, two have $(60,8,2)$, and one has $(66,0,4)$, where entries count deficits zero, one, and two. All ten matrices and full cut lists are supplied. The different profiles demonstrate that the objective can assign the same optimum to different entanglement distributions; neither the number of maximally mixed reductions nor mean entropy alone determines this cost.

The paper's previously obtained $(8,6)$ certificate has sector costs 28 and 16 and thus already attains $\mathcal C_6^*(8)=44$. Recomputing its complete cut profile gives 36 deficient cuts out of 162, all balanced, giving the original baseline of 34 maximally mixed balanced reductions. The aligned certificates in Appendix 8 attain 48 at the same cost and 56 at cost 48. The two independent prime-sector minima certify its objective value without asserting unrestricted optimality or existence of an AME state.

\setcounter{subsection}{0}
\section*{Appendix 6: Small-system optima and entropy profiles}
\subsection{An exact positive optimum within the four-quhex product ansatz}
For $N=4$ and $d=6=2\cdot3$, Eq.~\eqref{eq:composite_cost} has a strictly positive minimum over the CRT-product quadratic-phase family. No four-qubit AME state exists \cite{higuchi2000entangled,gour2010all}. Hence, for any $\mathbb F_2$ sector matrix at least one two-versus-two cut has rank below two. Its complementary two-site subset has the same rank, so those two terms alone contribute at least $2$ to the objective. This argument gives $\mathcal C_6(\mathbf P)\geq2$ for every matrix pair in the stated ansatz.

\begin{figure}[tb]
\centering
\includegraphics[width=0.49\textwidth]{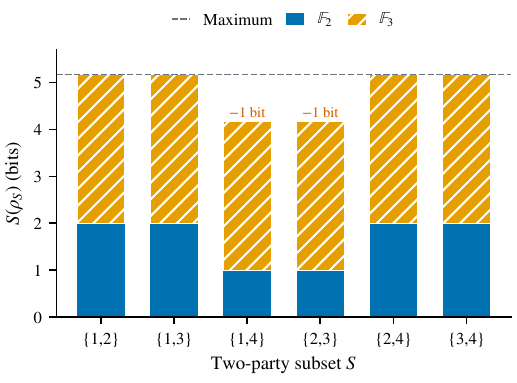}
\caption{Two-site entropies for the $N=4,d=6$ candidate in Eq.~\eqref{eq:four_six_matrices}, expressed in bits. The GF(3) sector reaches its two-site maximum on every cut. The GF(2) sector loses one bit on the complementary subsets $\{1,4\}$ and $\{2,3\}$, representing one balanced bipartition. The dashed line is $2\log_2 6$. The matrices are exactly optimal for the objective \eqref{eq:composite_cost} within the CRT-product quadratic-phase family. Unrestricted $\operatorname{AME}(4,6)$ states exist \cite{rather2022thirty,gross2025thirty}.}
\label{fig:near46}
\end{figure}

The bound is attained by the following matrices returned by AfAME:
\begin{equation}
\begin{aligned}
P^{(2)}&=
\begin{pmatrix}
0&1&1&0\\1&0&0&1\\1&0&0&1\\0&1&1&0
\end{pmatrix},\\
P^{(3)}&=
\begin{pmatrix}
0&1&0&2\\1&0&2&1\\0&2&0&2\\2&1&2&0
\end{pmatrix}.
\end{aligned}
\label{eq:four_six_matrices}
\end{equation}
Every one-site cut has full rank in both sectors. All six two-site cuts have rank two over $\mathbb F_3$; over $\mathbb F_2$ only $\{1,4\}$ and its complement $\{2,3\}$ have rank one. Direct substitution in Eq.~\eqref{eq:composite_cost} gives $\mathcal C_6=2$. Together with the lower bound and Eq.~\eqref{eq:min_separates}, this proves $\mathcal C_6^*(4)=\mathcal C_2^*(4)+\mathcal C_3^*(4)=2+0$ \emph{within the product ansatz}. The two deficient reduced states each have entropy $2\ln6-\ln2$; Fig.~\ref{fig:near46} shows their sector contributions.

\subsection{Recovery of the known seven-qubit extremum}
\begin{figure*}[tb]
\centering
\includegraphics[width=0.92\textwidth]{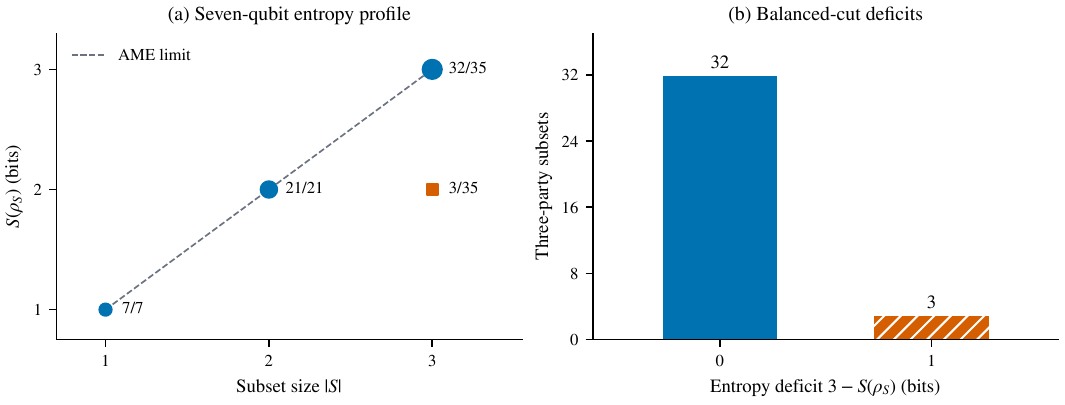}
\caption{Entropy profile and balanced-cut deficits of the seven-qubit candidate, in bits. Three of the $35$ three-qubit subsets have entropy $2$ rather than $3$ bits; every smaller subsystem attains its maximum. The cost $\mathcal C_2=3$ is optimal for the stated quadratic-phase objective by the known seven-qubit extremal bound \cite{huber2017absolutely}.}
\label{fig:near72}
\end{figure*}

For $N=7,d=2$, AfAME returned a two-uniform graph state with cost $\mathcal C_2=3$: all $7$ one-qubit and $21$ two-qubit cuts are full rank, whereas $32$ of $35$ three-qubit cuts have rank three and three have rank two. Figure~\ref{fig:near72} gives the complete entropy-deficit distribution. The independent verifier reproduces all $63$ cut ranks. Among two-uniform seven-qubit states, the upper bound of $32$ maximally mixed three-qubit reductions and an attaining state were already established by Huber \emph{et al.} \cite{huber2017absolutely}; the extremal question was revisited by Zhang \emph{et al.} \cite{zhang2025extremal}.

These prior results also show that $\mathcal C_2^*(7)=3$ within the quadratic-phase family. If every one- and two-site reduction is maximally mixed, at least three of the $35$ three-site reductions fail by the Huber \emph{et al.} bound, contributing at least $3$ to the integer cost. If a two-site reduction fails, each of its five three-site supersets fails because a maximally mixed three-site state would have maximally mixed two-site marginals; the cost is then at least $6$. If a one-site reduction fails, its six two-site supersets fail, so the cost again exceeds $3$. Our returned matrix attains the bound. This identifies an exact optimum of the stated objective, but the underlying extremal entanglement bound is established prior work.

\setcounter{subsection}{0}
\section*{Appendix 7: Composite search and matched-budget comparison}

AfAME evaluates changes under symmetric off-diagonal matrix updates, uses guided proposals with Metropolis--Hastings acceptance, and exchanges replicas at different temperatures \cite{afame2026software}. Restarts supply independent search opportunities. The final C++ cut-rank pass and a separate Python implementation recompute the exported certificate. This separates stochastic construction from exact verification. The code evaluates all subsets in $\mathcal S_N$; the cost of one rank calculation is polynomial in $N$, whereas the number of subsets is exponential.

The separability in Eq.~\eqref{eq:min_separates} also gives a deterministic improvement to the output of any search that visits a finite set $\mathcal T$ of composite matrix tuples. Let $c_\alpha(P^{(\alpha)})$ be sector $\alpha$'s contribution to Eq.~\eqref{eq:composite_cost}. Retain a sector matrix attaining $m_\alpha=\min_{t\in\mathcal T}c_\alpha(P_t^{(\alpha)})$, even when the minima occur at different steps, replicas, or restarts. The assembled tuple $\widehat{\mathbf P}$ then obeys
\begin{equation}
\mathcal C_d(\widehat{\mathbf P})
=\sum_\alpha m_\alpha
\leq\min_{t\in\mathcal T}\sum_\alpha c_\alpha(P_t^{(\alpha)}).
\label{eq:recombination}
\end{equation}
The inequality follows because every term of every visited tuple is at least its sector minimum. If every sector reaches zero somewhere along the search, their recombination is a zero-cost AME certificate within the product family even if the search never visits the combined tuple. The guarantee follows from established field-sector independence. The instrumented AfAME version supplied with the ancillary data stores the sector records and their source locations and rebuilds the cost cache before exporting the combined certificate.

\begin{table*}[t]
\caption{Held-out paired effect of sector-wise recombination at a fixed budget of 2000 local proposals per run. The intervals are percentile seed-bootstrap intervals for the mean paired cost reduction.}
\label{tab:recombination}
\centering
\begin{tabular}{c r r r r}
\toprule
$N$ & Joint mean & Recombined mean & Strict gains & Mean paired gain (95\% CI) \\
\midrule
10 & 200.98 & 200.58 & 9/50 & 0.40 (0.16, 0.68) \\
11 & 95.06 & 93.70 & 18/50 & 1.36 (0.80, 1.98) \\
12 & 867.60 & 858.18 & 34/50 & 9.42 (6.66, 12.40) \\
\bottomrule
\end{tabular}
\end{table*}

\begin{figure*}[t]
\centering
\includegraphics[width=0.92\textwidth]{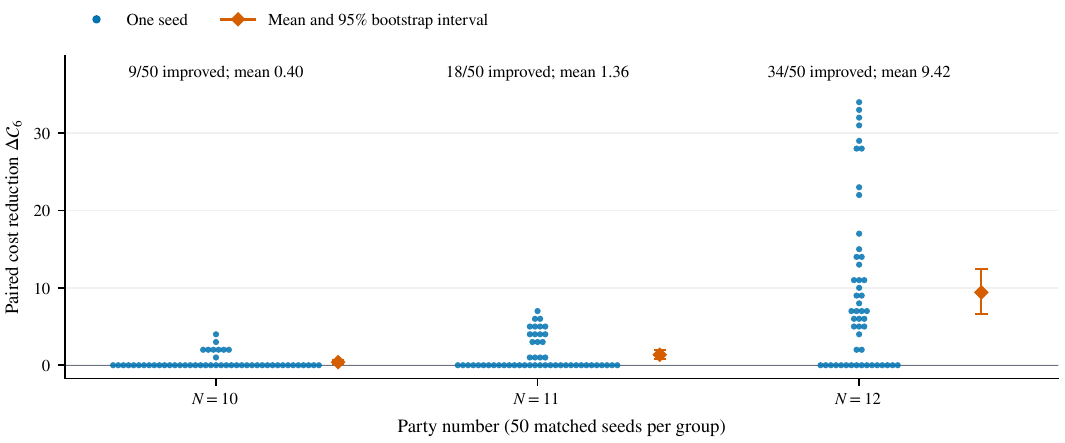}
\caption{Held-out effect of recombining the best GF(2) and GF(3) sector matrices after an unchanged AfAME trajectory. All 150 paired outcomes use a common vertical scale. Each blue dot represents one seed's reduction in cost at the same 2000-proposal budget; horizontal stacking separates equal values and has no quantitative meaning. Orange diamonds and whiskers show the mean and its 95\% percentile bootstrap interval, respectively. All zero differences are retained. Group labels give strict gains and mean reductions. The $N=12$ mean gain is 1.09\% of its joint-search mean; no runtime advantage is inferred.}
\label{fig:recombination}
\end{figure*}

We tested the gain at a fixed trajectory budget for $d=6$ with held-out seeds 1000--1049 at each of $N=10,11,12$. Each run used four replicas, 500 steps per replica and one restart, at most 2000 local proposals; all 150 runs made exactly 2000 proposals. The baseline is the best \emph{jointly visited} tuple on the same run, which is the cost returned by the previous export rule. The updated bookkeeping consumes no random draws. Identical-seed baseline reruns on nine sampled cases reproduced the recorded joint costs. Independent Python rank checks reproduced all 150 exported costs. The paired outcomes are summarized in Table~\ref{tab:recombination}.

\noindent\begin{minipage}{\linewidth}
The primary $N=12$ mean reduction is 1.09\% of the joint-search mean. The improvement is guaranteed for a \emph{fixed visited set}, and these data measure its size for the specified search settings. They do not establish a faster time to an AME certificate or superiority to independently optimized field sectors or other search algorithms. Figure~\ref{fig:recombination} displays all paired differences.
\end{minipage}

\setcounter{subsection}{0}
\section*{Appendix 8: Trade-off certificates and the eleven-qutrit witness}
\subsection{Independent checks and exact comparison rules}
The supplied profile file \texttt{data/ternary\_class\_profiles.json} contains all 817 classified eight-qutrit profiles used for the Pareto comparison. A vector is removed precisely when another has all four rank-tail counts at least as large and one strictly larger. The surviving vectors are Eq.~\eqref{eq:tail_frontier}, attained by records 741--745, 748, and 750 for the first vector; 746 and 747 for the second; and 749 for the third. The complete profiles, not only the surviving points, accompany the source bundle.

Two additional binary implementations evaluate the complete cost and balanced good-cut count of all 133,632 atlas extensions. One uses binary pivot elimination and the other enumerates every row combination. Both verify the boundary $(28,48),(32,56)$, including the absence of any state with cost below 32 and more than 48 full-rank balanced cuts. Every cost-28 extension has balanced profile $(48,20,2)$. This is a new aggregation of established graph-classification data, not a new graph classification. The known rank-table profile of Cabello \emph{et al.} supplies a direct cross-check of the minimum \cite{cabello2009eight,cabello2010erratum}.

For each new witness, finite-field elimination is compared on every scored subset with a separate C++ enumeration of labels $(x,Px)$ and supported stabilizer counts. The comparison includes all 1023 cuts of the eleven-qutrit state. Its support enumeration runs over exactly $3^{11}=177147$ labels. Product witnesses are checked separately in both prime sectors and then checked by the original AfAME certificate verifier. These computations verify arithmetic; published classification completeness and the analytical arguments in the main text supply the lower-bound coverage.

\subsection{Explicit aligned product states}
The ternary matrix used for both alignments is record 749 of the published length-eight database:
\begin{equation}
P_{749}=\begin{pmatrix}
0&1&1&1&2&1&2&1\\
1&0&1&1&1&2&2&1\\
1&1&0&1&0&0&1&2\\
1&1&1&0&0&0&0&0\\
2&1&0&0&0&0&1&1\\
1&2&0&0&0&0&1&1\\
2&2&1&0&1&1&0&0\\
1&1&2&0&1&1&0&0
\end{pmatrix}.
\end{equation}
Its balanced profile is $(66,0,4)$. We label rows and columns from 1 to 8. A permutation listed below maps old vertex $i$ to the $i$th listed new vertex, so $P'_{\pi(i),\pi(j)}=P_{ij}$.

For $(\mathcal C_6,F_6)=(44,48)$, use the binary witness of Appendix 4 and apply
\begin{equation}
 \pi_{44}=(1,2,3,4,5,8,6,7)
\end{equation}
to $P_{749}$. For $(\mathcal C_6,F_6)=(48,56)$, use the published $T_4$ binary graph \cite{zhang2025extremal},
\begin{equation}
P_{T_4}=\begin{pmatrix}
0&1&1&1&1&0&0&0\\
1&0&1&1&0&1&0&0\\
1&1&0&1&0&0&1&0\\
1&1&1&0&0&0&0&1\\
1&0&0&0&0&1&1&1\\
0&1&0&0&1&0&1&1\\
0&0&1&0&1&1&0&1\\
0&0&0&1&1&1&1&0
\end{pmatrix},
\end{equation}
and apply
\begin{equation}
 \pi_{48}=(1,2,3,8,4,7,5,6)
\end{equation}
to $P_{749}$. The machine-readable certificates contain the already permuted matrices and all field metadata, avoiding ambiguity about permutation direction. In both alignments the four deficient ternary subsets are contained in the deficient binary set. All 92 smaller subsets are maximally mixed in both sectors. These inclusion witnesses, together with the binary bounds, prove the count and Pareto claims without enumerating all relative sector permutations.

\subsection{Published eleven-qutrit matrix and its profile}
Danielsen explicitly prints the unique $\alpha=0$ graph-code representative in the length-eleven classification \cite{danielsen2012classification}. With the diagonal $\omega$ replaced by zero and off-diagonal minus signs interpreted as $2\in\mathbb F_3$, its adjacency matrix is
\begingroup\small\setlength{\arraycolsep}{3pt}
\begin{equation}
P_{11}=\begin{pmatrix}
0&1&0&2&1&0&2&1&0&0&0\\
1&0&1&0&2&2&1&0&0&0&0\\
0&1&0&1&2&1&0&0&0&1&0\\
2&0&1&0&2&2&0&2&0&2&2\\
1&2&2&2&0&2&2&0&0&0&1\\
0&2&1&2&2&0&2&2&2&1&0\\
2&1&0&0&2&2&0&0&2&0&1\\
1&0&0&2&0&2&0&0&2&2&2\\
0&0&0&0&0&2&2&2&0&1&1\\
0&0&1&2&0&1&0&2&1&0&0\\
0&0&0&2&1&0&1&2&1&0&0
\end{pmatrix}.
\label{eq:published_eleven_matrix}
\end{equation}
\endgroup
The independent support enumeration gives
\begin{equation}
W(1,y)=1+12y^5+888y^6+3960y^7+14970y^8
+42500y^9+66240y^{10}+48576y^{11}.
\end{equation}
The coefficients sum to $3^{11}$. The retrieved formula for $W_{11,\alpha}$ in Ref.~\cite{danielsen2012classification} prints 48756 as the constant coefficient of $y^{11}$; this differs from the directly enumerated 48576 and fails normalization by 180 at $\alpha=0$. Our lower bound uses only its weight-five coefficient, allowed $\alpha$ values, and uniqueness statement; the witness and its full enumerator are checked independently. We preserve this discrepancy in the provenance notes rather than silently transcribing the inconsistent last coefficient.

Exactly six five-subsets have deficit one:
\begin{equation}
\begin{split}
 &\{1,2,3,4,5\},\quad\{1,2,3,6,10\},\\
 &\{1,2,4,8,9\},\quad\{1,2,5,7,11\},\\
 &\{1,2,6,7,8\},\quad\{1,2,9,10,11\}.
\end{split}
\end{equation}
All other five-subsets and all smaller scored subsets have deficit zero. Parties 1 and 2 lie in all six bad subsets; every other party lies in two. For a fixed reference, these incidences are exactly the failing four-erasure patterns because every four-party physical marginal is maximally mixed. Averaging over all references gives 30 failures among $11\binom{10}{4}=2310$ labelled reference--erasure choices, hence the exact recovery fraction $76/77$. Each failure has mutual information $\ln3$, yielding Eq.~\eqref{eq:leakage_bound}. The lower-bound proof does not regenerate Danielsen's 56,005,876 distance-five classes. It uses their published exhaustive weight-five list, then excludes every lower-distance state by Eq.~\eqref{eq:uniformity_bound}. Uniqueness refers to Clifford equivalence and permutations; no uniqueness under unrestricted local unitaries, or of every state minimizing average leakage, is claimed.

\end{document}